\documentclass[twocolumn,preprintnumbers,amsmath,amssymb]{revtex4}

\usepackage{graphicx}
\usepackage{amssymb,amsmath}

\begin{document}

\title{
Entanglement entropy of integer Quantum Hall states in polygonal domains
}

\author{Iv\'an D. Rodr\'{\i}guez$^\dag$ and Germ\'an Sierra$^*$}

\affiliation{$^\dag$ National University of Maynooth, Dublin, Ireland, \\
$^*$ Instituto de F\'{\i}sica Te\'orica, UAM-CSIC, Madrid, Spain .}

\bigskip\bigskip\bigskip\bigskip


%


\begin{abstract}
The entanglement entropy of the integer Quantum Hall states
satisfies the area law for smooth domains with a vanishing topological term.
In this paper we consider polygonal domains for which the area law acquires a
constant term
that only depends on the angles of the vertices and we give a general expression for it.
We study also  the  dependence of the entanglement spectrum on the geometry and give it  a simple
physical interpretation.
\end{abstract}

\maketitle

\vskip 0.2cm


\def\beq{\begin{equation}}
\def\eeq{\end{equation}}
\def\barray{\begin{eqnarray}}
\def\earray{\end{eqnarray}}

\def\sc{\scriptsize}
\def\und{\underline}
\def\ov{\overline}
\def\I{{\rm Im}}
\def\R{{\rm Re}}
\def\Z{\mathbb{Z}}
\def\C{\mathbb{C}}
\def\RR{\mathbb{R}}
\def\nl{\nonumber \\}

\def\ww{\wedge}
\def\ra{\rangle}
\def\rra{\rangle\!\rangle}
\def\la{\langle}
\def\de{\partial}
\def\wt{\widetilde}
\def\wh{\widehat}
\def\Tr{{\rm Tr}}
\def\dag{\dagger}

\def\a{\alpha}
\def\b{\beta}
\def\g{\gamma}
\def\G{\Gamma}
\def\D{\Delta}
\def\d{\delta}
\def\e{\epsilon}
\def\eps{\varepsilon}
\def\z{\zeta}
\def\h{\eta}
\def\th{\theta}
\def\k{\kappa}
\def\l{\lambda}
\def\L{\Lambda}
\def\m{\mu}
\def\n{\nu}
\def\x{\xi}
\def\X{\Xi}
\def\p{\pi}
\def\P{\Pi}
\def\r{\rho}
\def\s{\sigma}
\def\S{\Sigma}
\def\t{\tau}
\def\f{\varphi}
\def\F{\Phi}
\def\c{\chi}
\def\w{\omega}
\def\W{\Omega}
\def\Th{\Theta}
\def\B{{\bf B}}
\def\ap{\approx}


\section{Introduction}

The area law satisfied by the entanglement entropy of the low energy states of
quantum many body systems in Condensed Matter and  Field Theory, has
become  one of the most fundamental tools to study the physical properties of these  complex systems \cite{amico,cirac,plenio}.
To define this entropy one considers a low energy state $\psi$, usually the ground
state, and computes the reduced density matrix $\rho_A$ by tracing
out the degrees of freedom outside a domain $A$.
The entanglement entropy $S_A(\psi)$, associated to the state $\psi$
and the domain $A$,  is defined as the von Neumann entropy
of the reduced density matrix $\rho_A$, i.e. $S_A(\psi) = - {\rm tr} \rho_A \log \rho_A$.
The area law states that $S_A(\psi)$ is proportional to the size of the boundary of $A$. In 3D this size is the area separating
$A$ from its environment, which gives  the law  its name.
In lower dimensions one should rather used the terms perimeter law in 2D  and  zeroth law in 1D, but those
names are not customary.

Three issues are important regarding the area law:   violations, fluctuations and
subleading corrections. They all provide a great amount of information about the system.
In conformal invariant 1D models,  the area law shows a log violation
proportional to  the central charge of the corresponding CFT and the topology open/close
of the system \cite{log,latorre,korepin,cardy}.
Fluctuations around  the log law for the Renyi entropy in Luttinger liquids allows a determination
of the Luttinger parameter \cite{cala-par},  and subleading terms  contain information about the scaling
dimensions of the operators \cite{cala-corre}.

In 2D much less is known about corrections to the area law.
In systems with topological order, the correction  is a constant term
$\gamma$ denoted  topological entanglement entropy \cite{kitaev,levin},
and whose value is given by the logarithm of the total quantum dimension of the anyonic excitations.
The area law of the Fractional Quantum Hall (FQH)  states
has been the target of several recent studies
\cite{schoutens1,schoutens2,laeuchli,friedman,fradkin}, in order
to confirm its validity and to compute
the value of $\gamma$ predicted in \cite{kitaev,levin}.
Reference \cite{fradkin} uses Chern-Simons theory, finding
the predicted value of
$\gamma$, however the linear behaviour of $S_A$,
is not captured,  due to the
purely topological nature of this theory. There are  numerical
studies using the Laughlin wave function \cite{schoutens1,schoutens2,laeuchli} and exact
diagonalization \cite{friedman,laeuchli}, for filling fractions
$\nu = 1/3, 1/5$ and the $\nu = 5/2$  Pfaffian state.
The approaches of \cite{schoutens1,schoutens2,laeuchli,friedman}
use the orbital basis for the Landau levels.
The close relationship of this basis to the spatial partioning
of the blocks leads to an area law of the form
$S_A = c \sqrt{l_A} - \gamma + O(1/l_A)$, where
$l_A$ is the number of Landau orbitals in the block $A$.
The numerical values of $\gamma$ computed in the spherical
geometry \cite{schoutens1} and the torus geometry \cite{friedman,laeuchli}  agree,
within some precision, with their theoretical values, despite
of the fact that the systems analyzed are not very large.

For non topological models there are also some results.
In those with  a Fermi surface,  the area law exhibits a log violation \cite{fermi1,fermi2,fermi3,fermi4,fermi5,fermi6} reminiscent of the 1D conformal systems, which suggest that
higher dimensional bosonization methods  may give an appropiate
description \cite{hal-boso}. Corrections to the area law in non-smooth domains, in $2+1$ dimensions, have been obtained  in relativistics free bosons and fermions in ref.\cite{CH} and in an interacting CFT using the Ads/CFT conjecture in \cite{Ads}.
The 2D quantum critical models of Fradkin and Moore \cite{FradMoore}  with critical exponent $z=2$ have also universal subleading
contributions  (see also \cite{fra1}).

The aim of this paper is to calculate the entanglement entropy of the IQHE ground state
in  arbitrary polygonal   domains.
We obtain that the entanglement entropy is given by the area (or perimeter) law plus a contribution due to the vertices, $\gamma$, that only depends on the angle of the vertices and the density of the fluid of electrons.

In the following we shall  concentrate in the IQHE state with filling fraction
$\nu=1$  defined on
a cylindrical geometry, but the calculation of $\gamma$ for higher filling fractions and different geometries are easily generalizable.

\begin{figure}[t!]
\begin{center}
\includegraphics[width= 9cm]{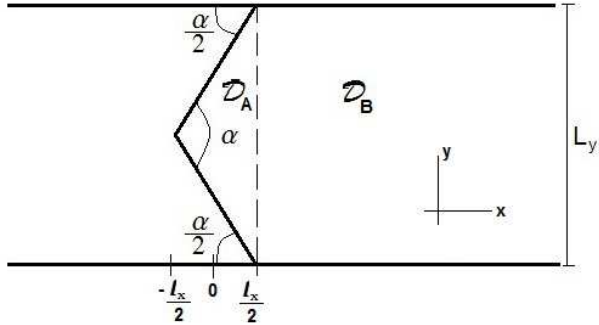}
\caption{Domain $\mathcal{D}=\mathcal{D}_A \cup \mathcal{D}_B$ including two vertices with angle $\alpha$. Varying the parameter $l_x=\frac{L Cot(\alpha/2)}{2}$ we can obtain the correction to the area law, $\gamma$, for $\alpha \in [0,\pi]$.}
\label{fig2}
\end{center}
\end{figure}

Let us consider the Landau model for a particle in a cylinder
of size $L_x \times L_y$. The one particle wave function in the lowest
Landau level (LLL),  in the gauge ${\bf A} = B(0,x)$,  is
(in units of the magnetic length $\ell $ equal to one):
\beq
\phi_{k_y}(x,y) = \frac{1}{ \pi^{1/4} L_y^{1/2}}  \; e^{i k_y y}
\; e^{- (x- k_y)^2/2} .
\label{1}
\eeq
On the cylinder, the identification of the wave function along the
$y$ direction implies:
\beq
k_y = \frac{2 \pi n}{L_y}, \quad   - \frac{n_0}{2} + 1 \leq n \leq
\frac{n_0}{2} .
\label{2}
\eeq
The number of LLLs, $n_0$, is obtained imposing that the particle
lives in the strip $|x| \leq L_x/2$, which yields
$n_0 = \frac{L_x L_y}{2 \pi}$.
This value also gives the total number of quantum fluxes
through the box.
The electron operator can be written as
\beq
\psi(x,y) = \sum_{k_y} \phi_{k_y}(x,y)  c_{k_y} + {\rm higher} \; {\rm LLs} ,
\label{6}
\eeq
where $c_{k_y}$ is the fermionic destruction operator of the LLL labeled by $k_y$.
The extra term in (\ref{6}) involves the remaining Landau levels, which are empty for filling fraction $\nu = 1$.
The ground state for $\nu = 1$ is given by:
\beq
|\Phi_0 \rangle = \Pi_{k_y} c_{k_y}^\dagger  |0 \rangle ,
\label{7}
\eeq
where $|0 \rangle $ is the Fock vacuum. The two point fermion correlator  in this state is,
\beq
C_{\textbf{r}, \textbf{r'}} = \langle \Phi_0 | \psi^\dagger(x,y)
\; \psi(x',y') | \Phi_0  \rangle .
\label{8}
\eeq
Using (\ref{6}) and (\ref{7}) one finds:
\beq
C_{\textbf{r}, \textbf{r'}} = \sum_{k_y}  \phi_{k_y}^*(x,y) \; \phi_{k_y}(x',y') .
\label{9}
\eeq
We want to  compute the entanglement entropy, $S_\mathcal{D}$,  of the
state $\Phi_0$, in a polygonal  domain $\mathcal{D}$  embedded in a cylinder of radius $L_y$ such as that shown in fig.\ref{fig2}.
This  entropy  is given by the formula $S_\mathcal{D} =
- {\rm tr}  \,  \rho_{\mathcal{D}} \log \rho_{\mathcal{D}} $ where
$\rho_{\mathcal{D}} =
{\rm tr}_{\mathcal{D}_c}
|\Phi_0 \rangle \langle \Phi_0|$, and  $\mathcal{D}_c$ is the complement
of $\mathcal{D}$ in the cylinder.  The computation of $S_\mathcal{D}$
is done in two steps \cite{latorre,peschel}. First one restricts the correlation matrix
$C_{\textbf{r}, \textbf{r'}}$, to the domain $\mathcal{D}$, i.e.
\beq
\tilde{C}_{\textbf{r}, \textbf{r'}} =  C_{\textbf{r}, \textbf{r'}}, \quad \textbf{r}, \textbf{r'} \in \mathcal{D} .
\label{13}
\eeq
Next, one diagonalizes $\tilde{C}_{\textbf{r}, \textbf{r'}}$, i.e.
\beq
\int_{\mathcal{D}} d^2  \textbf{r'} \; \tilde{C}_{\textbf{r}, \textbf{r'}} \; g(\textbf{r'} )
= \lambda g(\textbf{r}) .
\label{14}
\eeq
The  entropy $S_\mathcal{D}$ is  obtained by means of,
\beq
S_\mathcal{D} = \sum_m  H(\lambda_m) ,
\label{15}
\eeq
where $H(x) = - x \log x - (1-x) \log(1-x)$.
We have introduced a parameter $l_x$ that allow us to vary the angle $\alpha$ (see  fig.\ref{fig2}). If  $l_x=0$, the domain $\mathcal{D}(l_x=0)=\mathcal{D}_0$  becomes a  half cylinder. The eigenvalues of the correlator (\ref{13}) for this case were  given in  \cite{we}:
\barray
&& \lambda_n=\int_{\mathcal{D}_0} \phi^*_n(x,y) \phi_n(x,y) dx dy = \frac{1}{2} \left( 1 - Erf \left( \frac{2n\pi}{L_y} \right) \right) , \nl && n=-\infty,..,\infty \ .
\label{eigenlx0}
\earray
In this formula we assumed that $L_x$ is effectively infinite.
Computing now  the entanglement entropy (\ref{15}) using (\ref{eigenlx0}) one  obtains $S_{\mathcal{D}_0}=0.203291*L_y$. Observe that $L_y$ coincides with the perimeter of $\mathcal{D}_0$,  but this is a general result that holds for any smooth domain $\mathcal{D}_{smooth}$, i.e.:
\barray
S_{\mathcal{D}_{smooth}}= c*P ,
\label{smoothentropy}
\earray
where $P$ is the perimeter of $\mathcal{D}_{smooth}$ and the constant
\barray
c = 0.203291
\label{constant}
\earray
is independent of the geometry of the system but varies  with the number of  fully occupied  Landau levels \cite{we}.

\section{Correction to the area law}

The aim of this section is to show that the area law for a polygonal domain $\mathcal{D}$
is given by
%
\barray
S_\mathcal{D}= S_{\mathcal{D}_{smooth}} + \gamma ,
\label{entangvert}
\earray
where $S_{\mathcal{D}_{smooth}}$ is the entanglement entropy of a smooth domain $\mathcal{D}_{smooth}$ with the same perimeter as $\mathcal{D}$,
given by eq. (\ref{smoothentropy}),   and $\gamma$ is a constant term  due to the corners of the domain.

To proof eq.(\ref{entangvert}) we start by
diagonalizing  the correlator (\ref{13}) in a generic domain
$\mathcal{D}$.
Using (\ref{1}) and (\ref{9}) equation (\ref{14}) can be written as:
\barray
&& \sum_n \phi_n^*(x,y) \int_{D} \phi_n(x',y') g(x',y') =
\sum_n \phi_n^*(x,y) A_n \nl &=& \lambda g(x,y) \ ,
\label{densmat}
\earray
where we have defined:
\barray
A_n = \int_\mathcal{D} dx' dy' \phi_n(x',y') g(x',y') \ .
\label{definition}
\earray
The vanishing eigenvalues $\lambda$  of equation (\ref{14}),  does not  contribute to
$S_\mathcal{D}$, so one can  focus on the non vanishing ones. Using equation  (\ref{densmat}), the corresponding eigenfunctions
$g(x,y)$,  can be written as
\barray
g(x,y)= \frac{1}{\lambda} \sum_m \phi_m^*(x,y) A_m.
\label{gvsA}
\earray
Finally,  replacing (\ref{gvsA}) into (\ref{definition}) one obtains the following equation for the eigenvalues:
\barray
 \sum_m \left(F_\mathcal{D}\right)_{nm} A_m = \lambda A_n \ ,
\label{eigenprobl}
\earray
where $F_\mathcal{D}$ is a matrix with elements
\barray
\left(F_\mathcal{D}\right)_{nm} =
\int_\mathcal{D} \phi_n(x,y) \phi_m^*(x,y).
\label{eigenprobl2}
\earray
Observe that the continuous eigenvalue equation (\ref{14}) has been converted  into a discrete one (\ref{eigenprobl}),
where the eigenvalues of the matrix $F_{\mathcal{D}}$ are those of the correlator (\ref{13}). This fact will allow us to
apply numerical methods to find the eigenvalues $\lambda$.

\begin{figure}[t!]
\begin{center}
\includegraphics[width= 9cm]{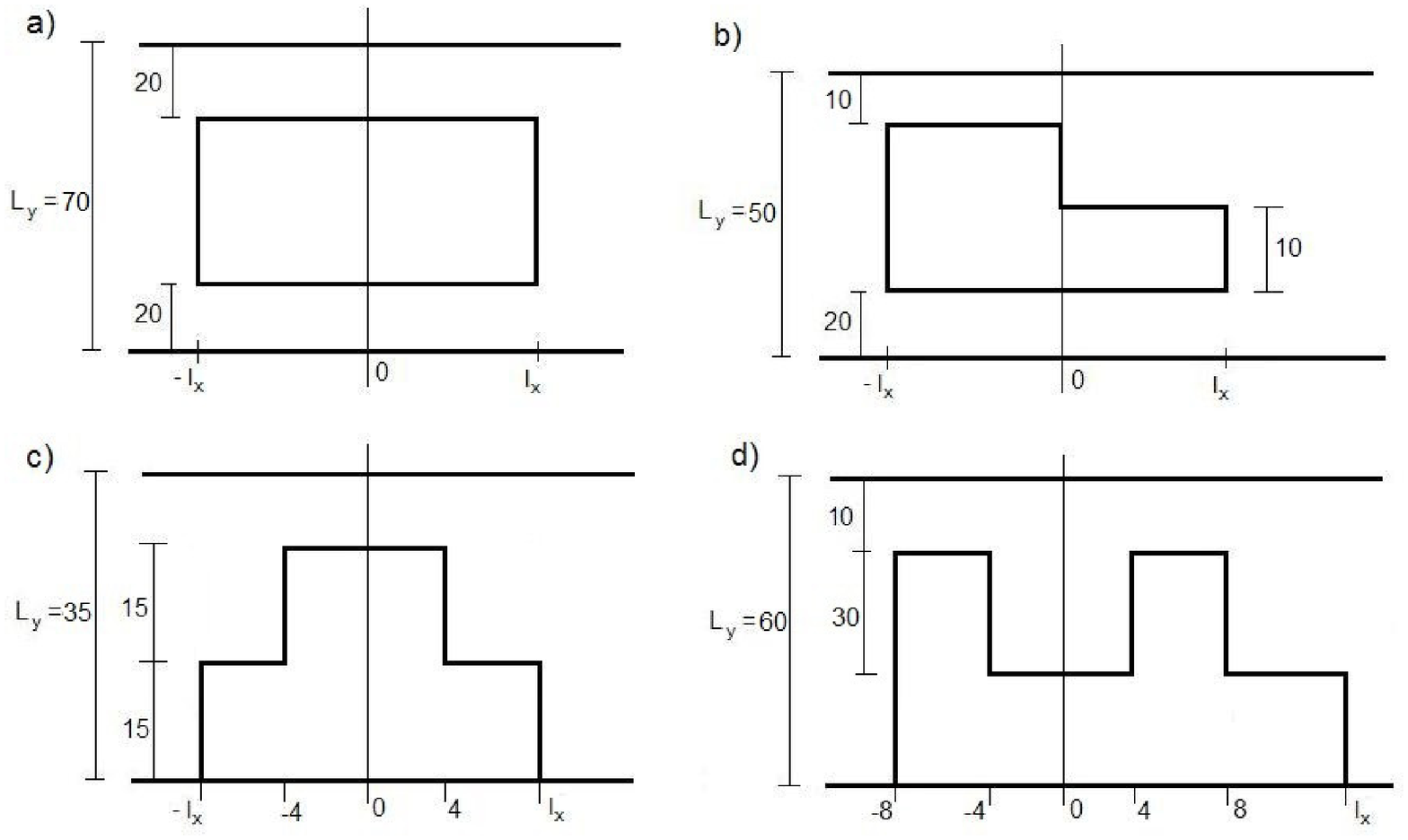}
\caption{Different domains in the torus with 4,6,8 and 10 right angles.}
\label{fig1}
\end{center}
\end{figure}

Indeed, let us first show the validity of equation (\ref{entangvert}) for  some simple domains as those depicted in
fig.\ref{fig1}.
 The domains $\mathcal{D}_n$, with $n=4,6,8,10$,
are  four different polygons with $n$ right angles.
For each of these domains we first compute the matrix $F_\mathcal{D}$  (see \ref{eigenprobl2}), and  we diagonalize it
for different values of $P$. The corresponding entropies $S_{\mathcal{D}_n}$ can be fitted to the formulas:
%

\vspace{0.3cm}

\noindent
a) $ S_{\mathcal{D}_4} = -0.341315 + 0.203291 * P_{\mathcal{D}_4} $, \\
b) $ S_{\mathcal{D}_6} = -0.512909 + 0.203295 * P_{\mathcal{D}_6} $, \\
c) $ S_{\mathcal{D}_8} = -0.683902 + 0.203291 * P_{\mathcal{D}_8} $, \\
d) $S_{\mathcal{D}_{10}} = -0.85412 + 0.203289 * P_{\mathcal{D}_{10}} $. \\ \\
Notice  that the constant multiplying the perimeters $P_{\mathcal{D}_n}$  agrees, to great accuracy,  with  the value $c$ in (\ref{smoothentropy})
for smooth domains.
%
%
%
 From these examples one can  extract the value of $\gamma$ for right angles:
\barray
\gamma \left( \frac{\pi}{2} \right) = a \left( \frac{\pi}{2} \right) *( \mbox{number of } \pi/2 \mbox{ vertices} )\ ,
\label{rightangles}
\earray
with $a\left( \frac{\pi}{2} \right)=-0.0855$. Based on (\ref{entangvert},\ref{rightangles}) we can propose
a general  expression for the entanglement entropy of an arbitrary domain $\mathcal{D}$ with $n$ vertices parameterized
by angles $\alpha_i \; (i=1, \dots, n)$
\barray
S_\mathcal{D} &=& S_{\mathcal{D}_{smooth}} + \gamma , \qquad \mbox{with} \nl
\gamma( \{ \alpha_i \} ) &= & \sum_i a (\alpha_i)* n_i \ ,  \quad \alpha_i \in [0, 2 \pi] , \nl
\label{generalgam}
\earray
with $n_i$ the number of vertices with angle
$\alpha_i$ in the boundary of $\mathcal{D}$. Each of these angles varies between 0 and $ 2 \pi$, but as we show below,
the parameter $\gamma$ is invariant under the symmetry $\alpha \leftrightarrow 2 \pi - \alpha$,
which follows from the equality of the entropy of a pure state in  a domain and its complement. For this reason we can
restrict $\alpha$  to the interval $\mbox{[} 0, \pi \mbox{]}$.

\begin{figure}[t!]
\begin{center}
\includegraphics[width= 7cm]{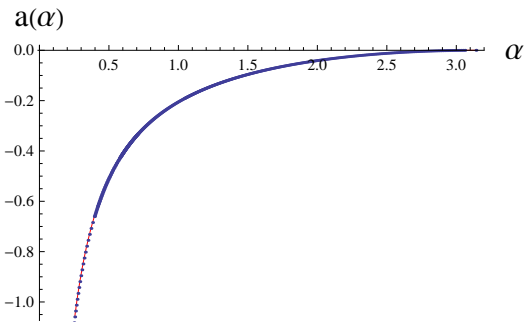}
\caption{Plot of $a(\alpha)$ vs. $\alpha$ for $\alpha \in [0, \pi ]$: the dotted
 line corresponds to the values of $a(\alpha)$ obtained by diagonalization of the $F_\mathcal{D}$ matrix (\ref{eigenprobl}) and
 the continuous line shows  fit  (\ref{fit2}). There is a perfect matching between the  two curves.}
\label{extrapol1}
\end{center}
\end{figure}

The calculation of $a(\alpha)$  can be done considering the domain of fig.\ref{fig2}. For different values of
$\alpha \in \mbox{[} 0, \pi \mbox{]}$ (or $l_x$ in fig.\ref{fig2}),  we diagonalize the $F_\mathcal{D}$ matrix,  calculate $S_\mathcal{D}$ and finally using (\ref{generalgam}),  we obtain the value of $a(\alpha)$ by means of the equation
\barray
a (\alpha)= \frac{1}{2} \left( S_\mathcal{D} - 0.20329*P_{\mathcal{D}} \right) \ ,
\label{aalfa}
\earray
with $P_{\mathcal{D}}$ the perimeter of $\mathcal{D}$. The  factor $\frac{1}{2}$ in (\ref{aalfa}) arises from the fact that
the domain of  fig.\ref{fig2} contains  two vertices with angle $\alpha$.
The numerical determination of  $a(\alpha)$ is given in fig.\ref{extrapol1}.
This curve can be fitted with the following expansion in the variable $ 2 l_x/L_y = \left|  \cot (\alpha/2) \right|  $ (see fig.\ref{fig2}) :
\barray
&& a(\alpha) =  0.00723706 \left| \cot (\alpha/2) \right|   - 0.155709 \left| \cot (\alpha/2) \right|  ^2 +  \nl &&  0.0876488 \left| \cot (\alpha/2) \right|  ^3 - 0.0306014 \left| \cot (\alpha/2) \right|  ^4 + \nl &&  0.00698332 \left| \cot (\alpha/2) \right|  ^5 - 0.00105786 \left| \cot (\alpha/2) \right|  ^6 + \nl && 0.000105322 \left| \cot (\alpha/2) \right|  ^7  -6.616*10^{-6} \left| \cot (\alpha/2) \right|^8 + \nl && 2.375*10^{-7} \left| \cot(\alpha/2)\right|^9  - 3.7112*10^{-9} \left| \cot(\alpha/2)\right|^{10} . \nl
\label{fit2}
\earray
Note that (\ref{fit2})
satisfies the relation $a(\alpha)=a(2\pi-\alpha)$ that together with (\ref{generalgam}) imply that $S_\mathcal{D}=S_{\mathcal{D}_c}$ ($\mathcal{D}_c$ is the complement of $\mathcal{D}$) as explained above. Hence, we can  restrict ourselves to the interval
$\alpha \in \mbox{[} 0, \pi \mbox{]}$ where $\cot(\alpha/2)\geq 0$ so that we can drop the absolute values in (\ref{fit2}).


\begin{figure}[t!]
\begin{center}
\includegraphics[width= 7.5cm]{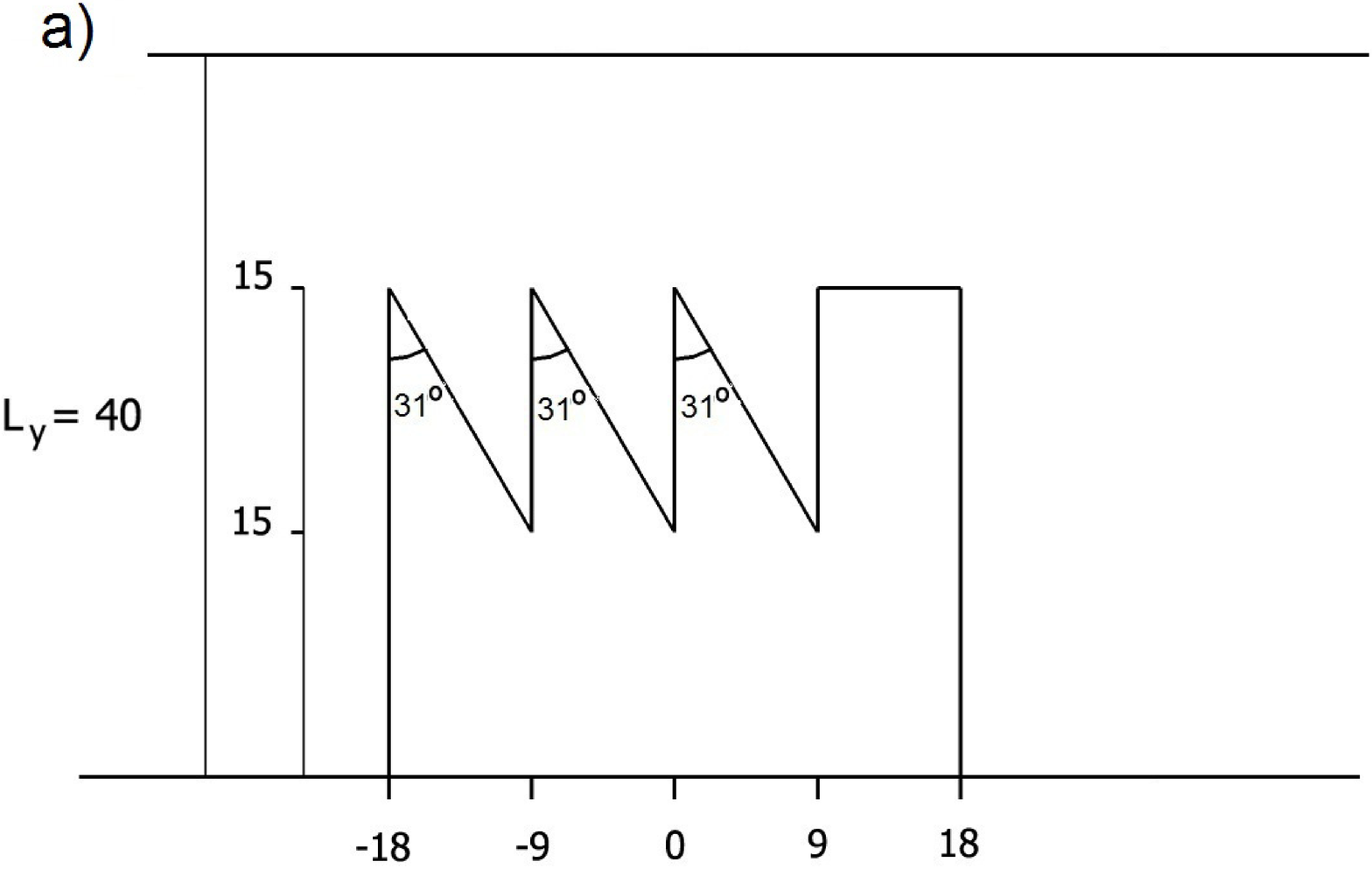}
\includegraphics[width= 8.5cm]{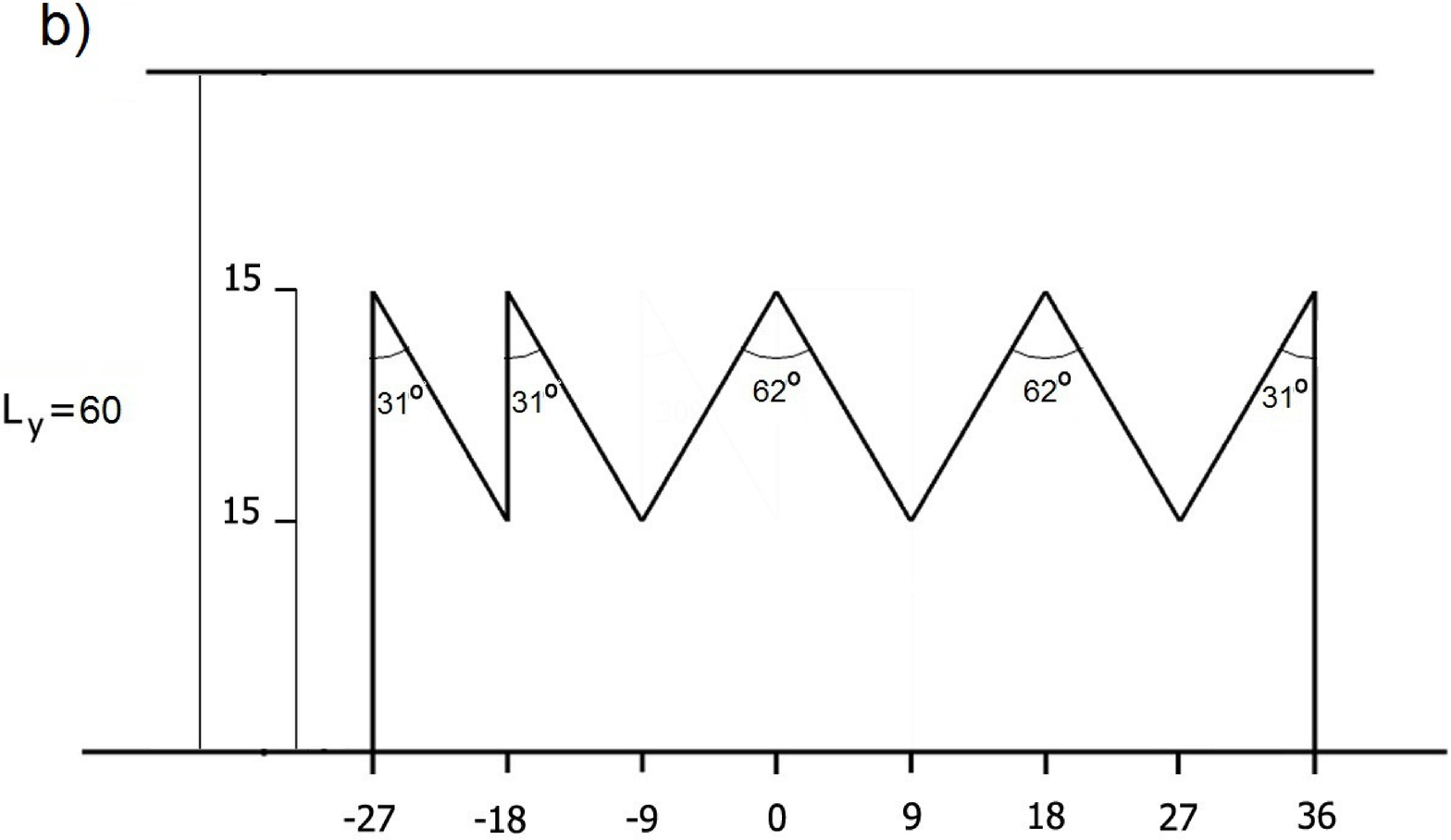}
\includegraphics[width= 8.5cm]{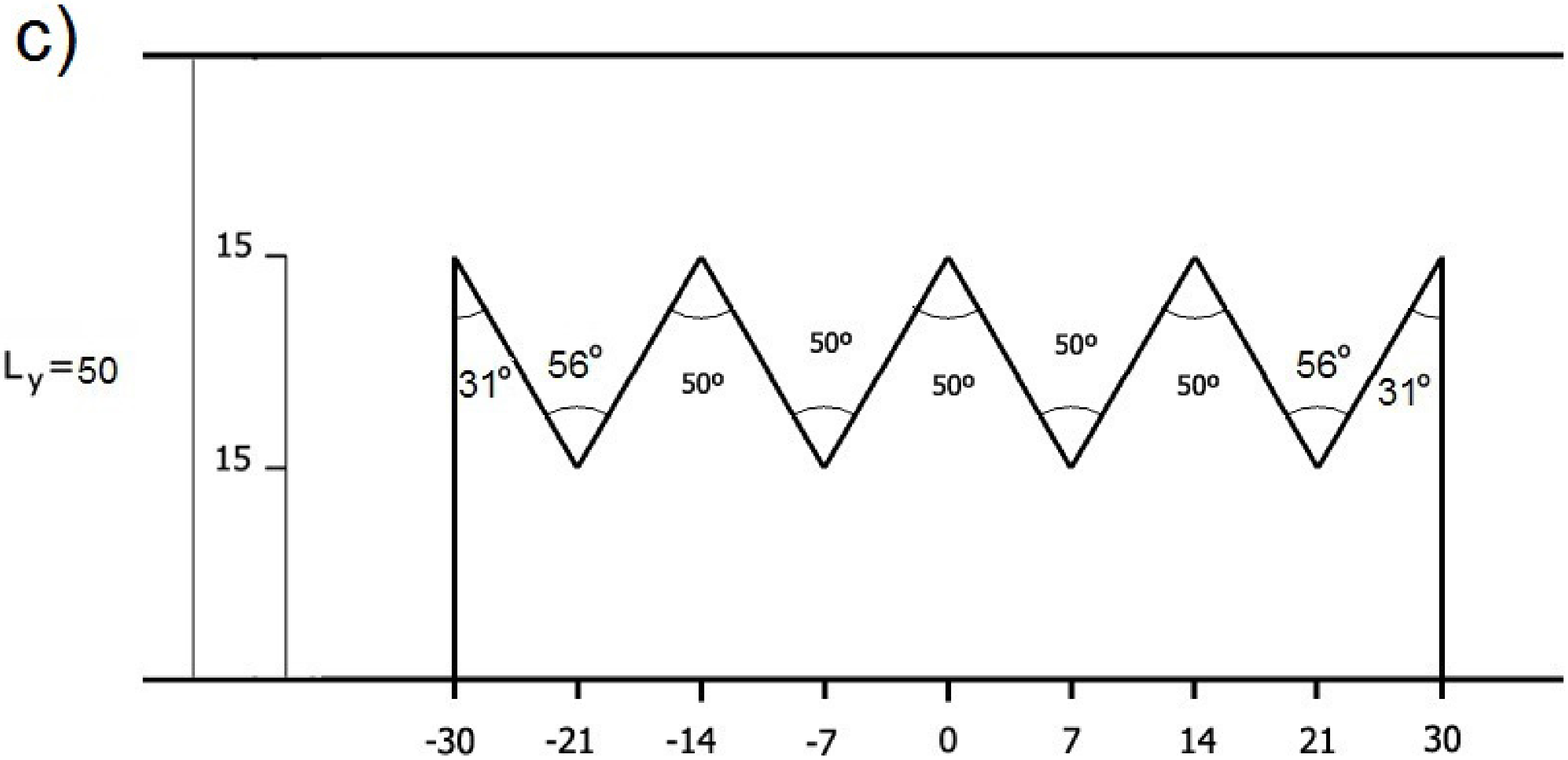}
\caption{The domains a) b) and c) with perimeters P=202.479, 260.45 and 254.3033 respectively are used to check the validity of (\ref{generalgam}) and (\ref{fit2}) . }
\label{fig4}
\end{center}
\end{figure}

In order  to assess the accuracy of
 the fit (\ref{fit2}) we can compare the entanglement entropies obtained by diagonalization of the $F_\mathcal{D}$ matrix (\ref{eigenprobl}), for the
 complicated  domains of  fig.\ref{fig4}, with the  theoretical formula (\ref{generalgam}) using the extrapolation (\ref{fit2}).
In the case of diagonalization of the $F_\mathcal{D}$ matrix we obtain:
\barray
S_a &=& 38.04348652 \ , \quad  S_b=50.02330873 \quad \mbox{and} \nl
S_c &=& 48.92750022 \
\label{cuenta1}
\earray
and using (\ref{generalgam}) and  (\ref{fit2}) we arrive at
\barray
S_a &=& 0.20329*P + 4*a(\pi/2) + 6*a(\arctan(9/15)) \nl &=& 38.047 , \nl
S_b &=& 0.20329*P + 2*a(\pi/2) + 4*a(\arctan(9/15)) + \nl && 5*a(2* \arctan(9/15))
= 50.0253 \qquad  \mbox{and} \nl
S_c &=& 0.20329*P + 2*a(\pi/2) + 2*a(\arctan(9/15)) + \nl
&& 2*a(\arctan(7/15) + \arctan(9/15)) + \nl && 5*a(2* \arctan(7/15))
= 48.927.
\label{cuenta2}
\earray
The comparison of  (\ref{cuenta1}) and (\ref{cuenta2}) shows clearly the validity of (\ref{generalgam}) and of the extrapolation (\ref{fit2}).
The previous results have been obtained for  a cylindrical geometry, however
 it can be shown that $a(\alpha)$ is the same function for the sphere and  the plane and thus $\gamma$ is also independent of the geometry.

\section{Analytical calculation of $a(\alpha)$.}

\begin{figure}[t!]
\begin{center}
\includegraphics[width= 7cm]{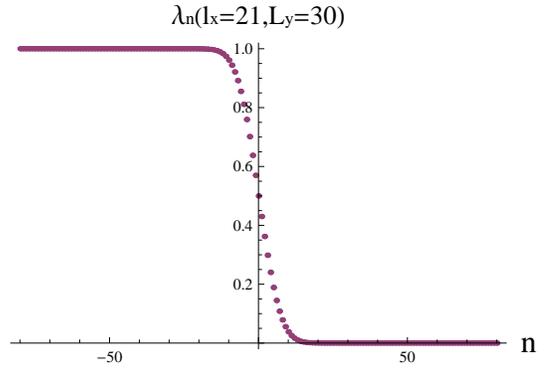}
\caption{Overlap between the eigenvalues $\lambda_n(l_x=21,L_y=30)$ obtained by diagonalization of the $F_\mathcal{D}$ matrix (\ref{eigenprobl}) and those obtained using the deformation (\ref{eigenvertex}) with $\kappa(l_x,L_y)=0.40266$.}
\label{fig5}
\end{center}
\end{figure}

In this section we shall  give an analytical justification of equation (\ref{generalgam}).
A crucial point in our analysis is  that the eigenvalue $\lambda_n(l_x,L_y)$ of the $F_\mathcal{D}$ matrix (\ref{eigenprobl}), corresponds to a deformation of the eigenvalue (\ref{eigenlx0}) for the case of the smooth domain $\mathcal{D}_0$ ($l_x=0$ in fig.\ref{fig2}).
This result   can be shown by diagonalization of $F_\mathcal{D}$ for many different values of $l_x$ and $L_y$. This deformation is parameterized by a function $\kappa(l_x,L_y)$ and it is given by
:
\barray
&& \lambda_n(l_x,L_y)=\frac{1}{2} \left( 1 - Erf \left[ \frac{2n\pi}{L_y} \left( 1-\kappa(l_x, L_y) \right) \right] \right).  \  \nl
\label{eigenvertex}
\earray
In fig.\ref{fig5} we show, for the particular case $l_x=21$ and $L_y=30$ ($\alpha=71.075^\circ$ in fig.\ref{fig5}), the perfect matching between the eigenvalues obtained by diagonalization of the $F_\mathcal{D}$ matrix with those given by (\ref{eigenvertex}).

To calculate the function $\kappa(l_x,L_y)$ we can use equations (\ref{15}),(\ref{generalgam}),(\ref{eigenvertex}) and the change of variables $x=(2 \pi n /L_y) \left( 1-\kappa(l_x,L_y) \right)$ obtaining:
\barray
&& S_{\mathcal{D}} = c P_{\mathcal{D}} + 2 a(\alpha) = \int_{-\infty}^{\infty} H[\lambda_n(l_x,L_y)] dn \nl &=& \frac{L_y c}{ (1-\kappa(l_x,L_y))} \ ,
\label{kapa}
\earray
with $P_{\mathcal{D}} = 2\sqrt{\left(\frac{L_y}{2}\right)^2 + l_x^2}$ the perimeter of $\mathcal{D}$ and $c$ given in (\ref{constant}). From equation (\ref{kapa}),
 the deformation $\kappa(l_x,L_y)$ is expressed by:
\barray
\kappa (l_x,L_y)= 1 - \frac{L_y c}{ (c P_{\mathcal{D}} + 2 a(\alpha))} , \
\label{kappa}
\earray
with $a(\alpha)$ given by the extrapolation (\ref{fit2}) or by the theoretical value (\ref{gammaanalytic}) to  be obtained below.

Equation (\ref{eigenvertex}) is rather interesting since it shows that the whole spectrum of the matrix $F_\mathcal{D}$ is given in terms, basically,  of the
function $\kappa (l_x,L_y)$. Hence,  we expect that overall quantities
like  $Tr \, F^p_{\mathcal{D}}=  \sum_n \lambda_n^p(l_x,L_y)$, will also depend on  the latter function. If for some $p$,  we were able to
compute this trace, then we will know $\kappa (l_x,L_y)$ as well $a(\alpha)$  by means of eq.(\ref{kappa}).
The simplest choice is $p=1$, but in this case the trace of $ F_\mathcal{D} $ does not depend on $\kappa (l_x,L_y)$.
This follows from the relations
\barray
Tr F_\mathcal{D} = \sum_n \lambda_n(l_x,L_y) = \int_\mathcal{D} C(x,y) dxdy \sim \rho_0 A ,
\label{ex}
\earray
where $C(x,y)$ is  the correlator (\ref{13}), $\rho_0$ is the density of electrons and $A >> 1 $ is  the area of the domain $\mathcal{D}$.

The next choice  is $p=2$ for which  $Tr \, F^2_{\mathcal{D}}$ does depend on $\kappa (l_x,L_y)$.
%
Indeed,  using (\ref{eigenvertex}) and taking the limit $n_0>>1$ we obtains:
\barray
&& Tr F^2_{\mathcal{D}} = \sum^{n0}_{n=-n0} \lambda_n^2(l_x,L_y) = \nl && n_0 +
\frac{L_y}{2 \sqrt{2} \pi^{3/2} \left( -1 + \kappa(l_x,L_y) \right)}
\ .
\label{Trazalamb}
\earray
Then from (\ref{Trazalamb}) and (\ref{kapa}), the entropy $S_\mathcal{D}$  and  $Tr  F^2_\mathcal{D} $, are related by
\barray
S_{\mathcal{D}}= - 2 \sqrt{2} \pi^{3/2} c \left( Tr F^2_{\mathcal{D}} - n_0 \right) \ .
\label{TrFEntro}
\earray
Therefore if we find an alternative formula for  $Tr  F^2_\mathcal{D} $,  we would  obtain a theoretical formula
for $S_\mathcal{D}$ and justify in this way equation (\ref{generalgam}). This is done
in the  appendix  for  the domain $\mathcal{D}$ of fig.\ref{fig2} with the result:
\barray
&& Tr F^2_{\mathcal{D}} =- \frac{P_{\mathcal{D}}}{2 \sqrt{2}
\pi^{3/2}} + n_0 \nl && + \frac{4+(\pi - \alpha)(-1+3Cos(\alpha))Csc(\alpha)}{4\pi^2}
\ ,
\label{analytTrF}
\earray
where  $P_\mathcal{D}$ is the perimeter of the domain $\mathcal{D}$.
Finally,   from (\ref{TrFEntro}) and (\ref{analytTrF}) we obtain:
\barray
S_\mathcal{D} &=& c P_\mathcal{D} + \gamma(\alpha)   \qquad \mbox{with} \nl \gamma(\alpha)&=& 2 a(\alpha) = -\frac{c}{\sqrt{2\pi}} \left(4+(\pi - \alpha)(-1+3 \cos(\alpha)) \times \right. \nl && \left. \csc(\alpha) \right) \ .
\label{gammaanalytic}
\earray
\begin{figure}[t!]
\begin{center}
\includegraphics[width= 7cm]{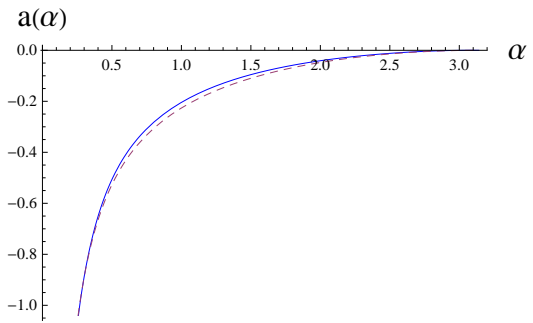}
\caption{Plot of $a(\alpha)$ vs. $\alpha$. Continuous curve corresponds to the extrapolation of $a(\alpha)$ eq.(\ref{fit2}) obtained by exact diagonalization of the $F_\mathcal{D}$ matrix. Dashed curve corresponds to the theoretical approximation for $a(\alpha)$ in eq.(\ref{gammaanalytic}).}
\label{fig6}
\end{center}
\end{figure}
Fig.(\ref{fig6}) shows  the overlap between the extrapolation $a(\alpha)$ in (\ref{fit2}) obtained by diagonalization of the $F_\mathcal{D}$ matrix and the theoretical expression (\ref{gammaanalytic}). The small difference between the curves is due to the approximation done in the calculation of $Tr  F_\mathcal{D}^2$ (see  eq.(\ref{intD}) in the Appendix).

In conclusion, in this section we have proof the validity of our proposal equation (\ref{generalgam}) for the entanglement entropy in a non-smooth domain.

\section{Entanglement spectrum}

The reduced density matrix $\mathcal{D}$ of a domain contains of course much more information that the entropy $S_\mathcal{D}$.
In the context of the QHE this information is related to the physical degrees of freedom of the edge excitations,  as proposed in reference
\cite{hald}. This is a rather surprising conjecture since, after all,  the edge of $\mathcal{D}$ is an arbitrary curve within the whole system.
The entanglement Hamiltonian, $H_\mathcal{D}$, defined as $\rho_\mathcal{D}=e^{-H_\mathcal{D}}$,  is then expected
to be intimately  related to the Hamiltonian describing the excitations of a real edge.  In the QHE the edge excitations
are described by a chiral CFT, which also describes the excitations of the bulk.
For the IQHE this result follows
from  the entanglement spectrum that is given  by the eigenvalues of the
 operator \cite{gd}
\barray
Q_\mathcal{D} = (1/2) - C_\mathcal{D} ,
\label{entspec}
\earray
with $C_\mathcal{D}$ the two point correlator defined in (\ref{13}). In the smooth domain $\mathcal{D}_0$ (obtained considering $l_x=0$ in fig.\ref{fig2}),  the eigenvalues of $C_{\mathcal{D}_0}$ are given by eq.(\ref{eigenlx0}),
so the entanglement spectrum will be:
\barray
&& q_n=
 \frac{1}{2} Erf\left( \frac{2 \pi n}{L_y} \right) \ , \nl
&& n=-\infty,...,\infty \ ,
\label{spectrumlx0}
\earray
with $q_n$ the eigenvalues of $Q_{\mathcal{D}_0}$.
In the limit $L_y>> 1$, this spectrum becomes $q_n = 2 \sqrt{\pi} n/L_y$, which is that  of a free chiral boson moving
on a circle of length $L_y$ with a velocity $v = 1/\sqrt{\pi}$. If we now deform  the domain as in fig. \ref{fig1},
  the entanglement spectrum in the boundary of $\mathcal{D}$ can be obtained from eq.
(\ref{eigenvertex})
\barray
&& q'_n=
\frac{1}{2} Erf\left( \frac{2 \pi n(1-\kappa(l_x,L_y)}{L_y} \right) \ , \nl
&& n=-\infty,...,\infty \ .
\label{spectrumlx}
\earray
In the limit $L_y >> 1$, eq.(\ref{kappa})  implies $P_\mathcal{D} = L_y/(1 - \kappa(l_x, L_y))$, and the spectrum (\ref{spectrumlx})
becomes  $q'_n = 2 \sqrt{\pi} n/P_\mathcal{D}$ which is that  of a free chiral boson moving
on a circle of length $P_\mathcal{D}$ with a velocity $v = 1/\sqrt{\pi}$. That circle has the same length  as  the boundary of $P_\mathcal{D}$.
The latter boundary has two singular points on it,  i.e. two vertices with angles $\alpha$ and $2 \pi - \alpha$, but they have subleading  effects both on the
entropy $S_\mathcal{D}$ and the entanglement spectrum. This phenomena is analogue to what happens for  real systems  where the
electrons surround the hills and valleys of the potential travelling all the way through the sample.

\vspace{1 cm}

In summary,  we have computed in this paper the subleading term   to the area law for  IQHE states on polygonal domains,
which is given by the expression

\barray
\gamma=\sum_i a(\alpha_i)*n_i
\label{gammaIQHE}
\earray
where $n_i$ is the number of vertices in the domain with angle $\alpha_i$ and $a(\alpha_i)$ is a function that only depends on the angles of the vertices and  the density of the fluid.  For the IQHE with filling fraction $\nu = 1$, we have found  numerical and analytical expressions of
 the function $a(\alpha)$, given by eqs.
 (\ref{fit2}) and (\ref{gammaanalytic}), which agree rather well.  The fact that the correction (\ref{gammaIQHE}) is a constant has its origin
in the gapped character of the IQHE. For gapless 2+1 systems one may expect a size dependent subleading  term.  Indeed,  this has been
confirmed for
 relativistics free bosons and fermions in ref.\cite{CH} and in an interacting CFT using the AdS/CFT conjecture in \cite{Ads}.
 Also, the critical models with dynamical exponent $z=2$ of Fradkin and Moore  \cite{FradMoore},  exhibit a logarithmic
 subleading correction of the form
\barray
\beta= f(\alpha)*Log(L)
\label{CFTcorrection}
\earray
with $L$ the perimeter of the domain and  $\alpha$ the angle of the vertex. The function $f(\alpha)$ has  a similar behavior as
$a(\alpha)$,  in the sense that both curves satisfy $f(\alpha), a(\alpha) \leq 0$ and $\frac{\partial f(\alpha)}{\partial \alpha}, \frac{\partial a(\alpha)}{\partial \alpha} \geq 0$. These properties are a consequence of the strong subadditivity relation satisfied by the entanglement entropy
\cite{Ads}.

Finally,  we have obtained the entanglement spectrum of  non smooth domains which corresponds to
a  chiral free boson moving on the boundary, in agreement with the conjecture of reference \cite{hald}.

~\\
 \noindent
{\it Acknowledgments}
We thank J.K. Slingerland and M. Haque for helpful comments.
This work has been supported by Science Foundation Ireland through PI Award 08/IN.1/I1961
(I. D. R.) and  the spanish  project FIS2009-11654 (G.S.). We  also acknowledge
ESF Science Programme INSTANS 2005-2010.

\subsection*{ Appendix}

In this Appendix we calculate the trace (\ref{Trazalamb})  analytically.
Let us first write it as
\barray
Tr F^2_{\mathcal{D}_A \cup \mathcal{D}_B} =Tr F^2_{\mathcal{D}_A}
+ Tr F^2_{\mathcal{D}_B} + 2 Tr \left(
F_{\mathcal{D}_A} F_{\mathcal{D}_B} \right) \ ,  \nl
\label{TrMatrix}
\earray
with $\mathcal{D}_A$ and $\mathcal{D}_B$ given in fig.\ref{fig2}.
From the definition of $F_\mathcal{D}$ (see eq.(\ref{eigenprobl})), the first term in (\ref{TrMatrix}) reads
\barray
&& Tr  F^2_{\mathcal{D}_A} =\int\int_{\mathcal{D}_A}d^2 r' \, d^2 r \sum^{n_0}_{m,n=-n_0} \bar{\phi}_m(x',y') \phi_n(x',y') \times \nl && \bar{\phi}_n(x,y) \phi_m(x,y) = \frac{1}{4\pi^2} \int \int_{\mathcal{D}_A} e^{-\frac{1}{2}(x-x')^2-\frac{1}{2}(y-y')^2} = \nl &&
\int^{\frac{l_x}{2}}_{-\frac{l_x}{2}} dx dx' \left(  \frac{e^{-\frac{1}{2}(x-x')^2 - \frac{1}{8} a^2 (l_x + x + x')^2}}{2 \pi^2} - \frac{e^{-\frac{1}{8}(4+a^2)(x-x')^2}}{2\pi^2} \right.  \nl && + \left.
\frac{ a e^{-\frac{1}{2}(x-x')^2}(x'-x)Erf\left[ \frac{a(x-x')}{2\sqrt{2}} \right]}{4\sqrt{2}\pi^{3/2}} +
\frac{a e^{-\frac{1}{2}(x-x')^2}}{4\sqrt{2}\pi^{3/2}} \times \right. \nl && \left. (l_x+x+x') Erf \left[ \frac{a(l_x+x+x')}{2\sqrt{2}} \right] \right) = A + B + C + D \ , \nl
\label{DADA}
\earray
where  $a=\frac{L_y}{l_x}$.
The integrals $A$ and $B$ in (\ref{DADA}) can be easily found
\barray
A &=& \frac{\pi + 2 \arctan\left[ \frac{1}{a} - \frac{a}{4} \right]}{4 \pi^2 a} \ , \nl
B &=& \frac{4}{(4+a^2)\pi^2} - \frac{\sqrt{2} \,  l_x}{\sqrt{4+a^2} \pi^{3/2}} \ .
\label{intAB}
\earray
To compute  the integral $C$ we first integrate the variables $y,y'$ and $x$ obtaining that:
\barray
&& C = \int_{-\frac{l_x}{2}}^{\frac{l_x}{2}} dx' \left( \frac{a}{2\sqrt{2}\pi^{3/2}} e^{-\frac{1}{8}(l_x-2x')^2} Erf \left[ \frac{a(l_x -2x')}{4\sqrt{2}} \right] \right. \nl && - \left. \frac{a^2}{2 \pi^{3/2} \sqrt{2(4+a^2)}} Erf \left[ \frac{\sqrt{4+a^2} (l_x-2x')}{4\sqrt{2}} \right] \right) = \nl && C_1 + C_2 \ .
\label{integralC}
\earray
The integral $C_1$ only depends on the ratio $a=L_y/l_x$. Therefore if we vary $l_x$, maintaining the value of $a$ constant (by adjusting $L_y$),  the integral $C_1$ remains the same. Taking  the limit $l_x \rightarrow \infty$ in $C_1$ and making  the change of variables $u=l_x-2x'$, one
obtains an integral between $0$ and $\infty$ whose value is
\barray
C_1=\frac{a}{2\pi^2} \arctan \left[ \frac{a}{2} \right] \ .
\label{intC1}
\earray

The integral $C_2$ is easily done
\barray
C_2 = \frac{a^2}{(4+a^2)\pi^2} - \frac{a^2 l_x}{2\sqrt{2(4+a^2)} \pi^{3/2}} \ .
\label{intC2}
\earray
Collecting the previous expressions one gets
\barray
C= \frac{a}{2\pi^2}  \arctan \left[ \frac{a}{2} \right] + \frac{a^2}{(4+a^2)\pi^2} - \frac{a^2 l_x}{2\sqrt{2(4+a^2)} \pi^{3/2}} \ . \nl
\label{intC}
\earray
The integral $D$ in (\ref{DADA}):
\barray
D=\int^{\frac{l_x}{2}}_{-\frac{l_x}{2}}\frac{a e^{-\frac{1}{2}(x-x')^2} (l_x+x+x') Erf \left[ \frac{a(l_x+x+x')}{2\sqrt{2}}\right]}{4 \sqrt{2} \pi^{3/2}} \ . \nl
\label{intD}
\earray
cannot be solved analytically. However,  one can obtain a very good approximation replacing in (\ref{intD}) $(l_x+x+x')Erf \left[ \frac{a(l_x+x+x')}{2\sqrt{2}}\right]$ by $(l_x+x+x')$, which upon integration yields
\barray
D=-\frac{a l_x}{2 \sqrt{2} \pi^{3/2}} + \frac{a l_x^2}{4 \pi} \ .
\label{iintD}
\earray
Finally,  from (\ref{intAB}),(\ref{intC}) and (\ref{iintD}) we arrive at:
\barray
&& Tr  F^2_{\mathcal{D}_A} = \frac{\pi + 2   \arctan\left[ \frac{1}{a} - \frac{a}{4} \right]}{4 \pi^2 a} \
+ \frac{4}{(4+a^2)\pi^2} - \nl && \frac{\sqrt{2}l_x}{\sqrt{4+a^2} \pi^{3/2}} + \frac{a}{2\pi^2}  \arctan \left[ \frac{a}{2} \right] + \frac{a^2}{(4+a^2)\pi^2} - \nl && \frac{a^2 l_x}{2\sqrt{2(4+a^2)} \pi^{3/2}} -\frac{a l_x}{2 \sqrt{2} \pi^{3/2}} + \frac{a l_x^2}{4 \pi}  \ .
\label{TrazaDA}
\earray
To complete the calculation one needs  the quantities $Tr  F^2_{\mathcal{D}_B} $ and
$2 Tr \left(F_{\mathcal{D}_A} F_{\mathcal{D}_B} \right)$ in (\ref{TrMatrix}). Their values are easy to obtain and they read
\barray
&& Tr F^2_{\mathcal{D}_B} = \sum_n \left( F_{\mathcal{D}_B} \right)_{nn} \left( F_{\mathcal{D}_B} \right)_{nn} =
n_0 - \frac{L}{2\sqrt{2}\pi^{3/2}} - \frac{L l_x}{4 \pi} \nl &&
Tr F_{(\mathcal{D}_A \mathcal{D}_B)}= \sum_n \left(F_{\mathcal{D}_A}\right)_{nn} \left(F_{\mathcal{D}_B}\right)_{nn} = \frac{L}{2\sqrt{2}\pi^{3/2}}-\frac{L}{8 l_x \pi}, \nl
\label{TraceDADB}
\earray
with $n_0$ defined in (\ref{2}). In (\ref{TraceDADB}) we have used the fact that in the domain $\mathcal{D}_B$,  the matrix $F$ is diagonal.

Finally,  from (\ref{TrazaDA}) and (\ref{TraceDADB}) we obtain that:
\barray
&&  Tr F^2_{\mathcal{D}_A \cup \mathcal{D}_B}=- \frac{P_{\mathcal{D}_A \cup \mathcal{D}_B}}{2 \sqrt{2} \pi^{3/2}} + n_0
+ \frac{4}{(4+a^2)\pi^2} + \frac{a^2}{(4+a^2)\pi^2} \nl && + \frac{1}{4 a \pi} - \frac{a}{4\pi} +
\frac{\arctan\left[\frac{1}{a} - \frac{a}{4} \right]}{2 a \pi^2} + \frac{a  \arctan \left[\frac{a}{2} \right]}{2 \pi^2}
\ ,
\label{TrazaFF}
\earray
with $P_{\mathcal{D}_A \cup \mathcal{D}_B}$ the perimeter of the domain $\mathcal{D}$ in fig.\ref{fig2}.
Observe from fig.\ref{fig2} that the ratio $a=\frac{L_y}{lx}$ is related with the angle, $\alpha$, of the vertex by
$a=2Tan(\alpha/2)$. Therefore we can write eq.(\ref{TrazaFF}) in terms of the angle $\alpha$ obtaining that:
\barray
&& Tr F^2_{\mathcal{D}_A \cup \mathcal{D}_B} =- \frac{P_{\mathcal{D}_A \cup \mathcal{D}_B}}{2 \sqrt{2}
\pi^{3/2}} + n_0 \nl && + \frac{4+(\pi - \alpha)(-1+3 \cos(\alpha)) \csc(\alpha)}{4\pi^2} \ .
\earray


\begin{thebibliography}{999}


\bibitem{amico} L. Amico, R. Fazio, A. Osterloh and V. Vedral, "Entanglement in Many-Body Systems",
Rev. Mod. Phys. vol. 80, 517-576 (2008); arXiv:quant-ph/0703044.

\bibitem{cirac} M.M. Wolf, F. Verstraete, M.B. Hastings and J.I. Cirac,
"Area laws in quantum systems: mutual information and correlations",
Phys. Rev. Lett. 100, 070502 (2008); arXiv:0704.3906.

\bibitem{plenio} J. Eisert, M. Cramer and M.B. Plenio,
"Area laws for the entanglement entropy - a review",
 Rev. Mod. Phys. 82, 277 (2010); arXiv:0808.3773.

\bibitem{log}  C. Holzhey, F. Larsen and F. Wilczek,
"Geometric and Renormalized Entropy in Conformal Field Theory",
  Nucl. Phys. B424, 443 (1994);  arXiv:hep-th/9403108.

\bibitem{latorre}  G. Vidal, J. I. Latorre, E. Rico and A. Kitaev,
"Entanglement in quantum critical phenomena",
Phys. Rev. Lett. 90, 227902 (2003);  arXiv:quant-ph/0211074.

\bibitem{korepin} B.-Q.Jin and V.E.Korepin,
"Quantum Spin Chain, Toeplitz Determinants and Fisher-Hartwig Conjecture",
 J. Stat. Phys. 116, Nos. 1-4, 79 (2004);  arXiv:quant-ph/0304108.

\bibitem{cardy} P. Calabrese and J. Cardy,
 "Entanglement Entropy and Quantum Field Theory",  J. Stat. Mech. (2004) P06002; arXiv:hep-th/0405152.

\bibitem{cala-par} P. Calabrese, M. Campostrini, F. Essler and  B. Nienhuis,
"Parity effects in the scaling of block entanglement in gapless spin chains",
 Phys.Rev.Lett.104, 095701 (2010); arXiv:0911.4660.

\bibitem{cala-corre} J. Cardy and  P. Calabrese,
"Unusual Corrections to Scaling in Entanglement Entropy",
J. Stat. Mech. (2010) P04023; arXiv:1002.4353.

\bibitem{kitaev}  A. Kitaev and J. Preskill,
"Topological entanglement entropy",
Phys. Rev. Lett. 96, 110404  (2006); arXiv:hep-th/0510092.

\bibitem{levin}   M. Levin and X. G. Wen,
"Detecting topological order in a ground state wave function",
Phys. Rev. Lett. 96, 110405 (2006);  arXiv:cond-mat/0510613.

\bibitem{schoutens1} M. Haque, O. Zozulya and  K. Schoutens,
"Entanglement entropy in fermionic Laughlin states",
Phys. Rev. Lett. 98, 060401 (2007); arXiv:cond-mat/0609263.

\bibitem{schoutens2} O.S. Zozulya, M. Haque, K. Schoutens and E.H. Rezayi,
"Bipartite entanglement entropy in fractional quantum Hall states"
Phys.  Rev. B76, 125310 (2007);  arXiv:0705.4176.

\bibitem{laeuchli} A. Laeuchli, E. J. Bergholtz and M. Haque,
"Entanglement Scaling of Fractional Quantum Hall states through Geometric Deformations";
arXiv:1003.5656.

\bibitem{friedman} B. A. Friedman and G. C. Levine,
"Topological entropy of realistic quantum Hall wave functions",
 Phys. Rev. B 78,035320 (2008); arXiv:0710.4071.

\bibitem{fradkin} S. Dong, E. Fradkin, R. G. Leigh and S. Nowling,
"Topological Entanglement Entropy in Chern-Simons Theories and Quantum Hall Fluids",
JHEP 0805: 016 (2008); arXiv:0802.3231.

\bibitem{fermi1} T. Barthel, M.C. Chung and U. Schollwock,
"Entanglement scaling in critical two-dimensional fermionic and bosonic systems",
Phys. Rev. A 74, 022329 (2006);arXiv:cond-mat/0602077.

\bibitem{fermi2} M. Cramer, J. Eisert and M.B. Plenio,
"Statistics dependence of the entanglement entropy",
Phys.Rev.Lett.98:220603 (2007); arXiv:quant-ph/0611264v4.

\bibitem{fermi3} S. Farkas and Z. Zimboras,
"The von Neumann entropy asymptotics in multidimensional fermionic systems",
J. Math. Phys. 48, 102110 (2007); arXiv:0706.1805v1.

\bibitem{fermi4} D. Gioev and I. Klich,
"Entanglement entropy of fermions in any dimension and the Widom conjecture",
Phys. Rev. Lett. 96, 100503 (2006);  arXiv:quant-ph/0504151.

\bibitem{fermi5} W. Li, L. Ding, R. Yu, T. Roscilde and S. Haas,
"Scaling Behavior of Entanglement in Two- and Three-Dimensional Free Fermions",
Phys.Rev. B 73, 064406 (2006); arXiv:quant-ph/0602094.

\bibitem{fermi6} M. M. Wolf,
"Violation of the entropic area law for Fermions",
Phys. Rev. Lett. 96, 010404 (2006); arXiv:quant-ph/0503219.

\bibitem{hal-boso}  F. D. M. Haldane,
"Luttinger's Theorem and Bosonization of the Fermi Surface",
in Perspectives in Many-Particle Physics, eds. R. Broglia and
J. R. Schrieffer, (North Holland, Amsterdam 1994, pp 5-30); cond-mat/0505529.

\bibitem{CH} H.Casini and  M.Huerta,
"Entanglement entropy in free quantum field theory",
J.Phys.A42:504007 (2009); airxiv:0905.2562v3 and
H. Casini, M. Huerta and  L. Leitao,
"Entanglement entropy for a Dirac fermion in three dimensions: vertex contribution",
Nucl.Phys.B814:594-609 (2009); arxiv:hep-th/0811.1968.

\bibitem{Ads}
T. Hirata and T. Takayanagi,
"AdS/CFT and Strong Subadditivity of Entanglement Entropy",  JHEP 0702, 042 (2007);
 arXiv:hep-th/0608213.

\bibitem{FradMoore}
E.  Fradkin and  J. E. Moore,
"Entanglement entropy of 2D conformal quantum critical points: hearing the shape of a quantum drum",
Phys.Rev.Lett.97, 050404 (2006);  airxiv:cond-mat/0605683.

\bibitem{fra1} B. Hsu, M.  Mulligan, E. Fradkin, and E. A. Kim,
"Universal entanglement entropy in 2D conformal quantum critical points",
Phys. Rev. B, 79, 115421 (2009); arXiv:0812.0203;
E. Fradkin,
" Scaling of Entanglement Entropy at 2D quantum Lifshitz fixed points and topological fluids",
Journal of Physics A: Math. Theor. 42, 504011 (2009); arXiv:0906.1569;
and  B. Hsu and E. Fradkin,
"Universal Behavior of Entanglement in 2D Quantum Critical Dimer Models"; arXiv:1006.1361.

\bibitem{peschel}
I. Peschel,
"Calculation of reduced density matrices from correlation functions",
J.Phys. A: Math. Gen. 36, L205 (2003);  arXiv:cond-mat/0212631.

\bibitem{we} Ivan D. Rodriguez and German Sierra,
"Entanglement entropy of integer Quantum Hall states",
Phys.Rev. B vol. 80, 153303 (2009); arXiv:0811.2188.

\bibitem{hald}  H. Li and F. D. M. Haldane,
"Entanglement Spectrum as a Generalization of Entanglement Entropy:
 Identification of Topological Order in Non-Abelian Fractional Quantum Hall Effect States",
Phys. Rev. Lett. 101, 010504 (2008); arXiv:0805.0332.

\bibitem{gd} M.A. Turner, Y. Zhang and A. Vishwanath,
 "Band Topology of Insulators via the Entanglement Spectrum";
 arXiv:0909.3119.

\bibitem{thomale} R. Thomale, A. Sterdyniak, N. Regnault and B. A. Bernevig,
"The entanglement gap and a new principle of adiabatic continuity",
Phys. Rev. Lett. 104, 180502 (2010); arXiv:0912.0523 and A. Sterdyniak, N. Regnault and B.A. Bernevig,
"Extracting Excitations From Groundstate Entanglement"; arXiv:1006.5435.

\end{thebibliography}
\end{document}